\documentclass[proceedings]{JHEP3}
\usepackage{amsfonts}
\usepackage{amsmath}
\usepackage{epsfig,multicol}

\newbox\mybox

\newcommand\fverb{\setbox\mybox=\hbox\bgroup\verb}
\newcommand\fverbdo{\egroup\medskip\noindent\fbox{\unhbox\mybox}\ }
\newcommand\fverbit{\egroup\item[\fbox{\unhbox\mybox}]}
\conference{From real fields to complex Calogero particles}
\abstract{We provide a novel procedure to obtain complex
$\mathcal{PT}$-symmetric multi-particle Calogero systems. Instead of
extending or deforming real Calogero systems, we explore here
the possibilities for complex systems to arise
from real nonlinear field equations. We exemplify this procedure
for the
Boussinesq equation and demonstrate how singularities in real valued wave solutions can be interpreted as $N$ complex particles scattering amongst each other. We analyze this phenomenon in more detail for the two and three particle case. Particular attention is paid to the implemention of $\mathcal{PT}$-symmetry for the complex  multi-particle systems. New complex $\mathcal{PT}$-symmetric Calogero systems together with their classical solutions are derived.}

\title{From real fields to complex Calogero particles}
\author{Paulo E. G. Assis and Andreas Fring \\
Centre for Mathematical Science, City University London, \\
Northampton Square, London EC1V 0HB, UK \\
E-mail: Paulo.Goncalves-de-Assis.1@city.ac.uk, A.Fring@city.ac.uk}

\input{tcilatex}

\begin{document}

\section{Introduction}

\label{secintro} \setcounter{equation}{0}

The analytic continuation of real physical systems into the complex plane is
a principle which has turned out to be very fruitful, since many new
features can be revealed in this manner which might otherwise be undetected.
A famous and already classical example, proposed more than half a century
ago, is for instance Heisenberg's programme of the analytic S-matrix \cite%
{heis}. Here our main concern will be complex multi-particle Calogero
systems, in particular those exhibiting $\mathcal{PT}$-symmetry \cite%
{benderb}.

Quantum systems are said to be $\mathcal{PT}$-symmetric when they are
invariant under simultaneous parity $\mathcal{P}$ and time reversal $%
\mathcal{T}$ transformations. When the Hamiltonian, not necessarily
Hermitian, exhibits this symmetry, i.e.~$[H,\mathcal{PT}]=0$, and moreover
when all wave-functions are also invariant under such an operation this
property is referred to as unbroken $\mathcal{PT}$-symmetry. The virtue of
this feature is that it is a sufficient property to guarantee the spectrum
of the Hamiltonian $H$ to be real. The underlying mechanisms responsible for
this are by now well understood \cite{wigner, ffaria, bender, mostaf} and
may be formulated alternatively in terms of pseudo/quasi-Hermiticity; for
definitions see for instance \cite{aflie} and references therein.

There are two fundamentally different possibilities to view complex systems:
One may either regard the complexified version just as a broader framework,
as in the spirit of the analytic S-matrix, and restrict to the real case in
order to describe the underlying physics or alternatively one may try to
give a direct physical meaning to the complex models.

With the latter motivation in mind complex $\mathcal{PT}$-symmetric Calogero
systems have been introduced and studied recently \cite{basu, taterz,
calfring, jain,fringz}. The hope for a direct physical interpretation stems
from the fact that unbroken $\mathcal{PT}$-symmetry will guarantee their
eigenspectra to be real and allows for a consistent quantum mechanical
description, i.e.~such systems constitute well defined quantum systems which
have been overlooked up to now. Nonetheless, so far any such proposal lacks
a direct physical meaning and the complexifications are generally introduced
in a rather ad hoc manner. Here our main purpose is to demonstrate that
various complex Calogero models appear rather naturally from \textit{real}
valued nonlinear field equations and thus we provide a well defined physical
origin for these systems.

The solutions for the real Calogero systems were found in the reverse order
when compared to the usual way progress is made, i.e.~the quantum theory was
solved before the classical one. Calogero solved first the quantized
one-dimensional three-body problem with pairwise inverse square interaction
\cite{calo3} and subsequently constructed the ground state of the $N$-body
generalization \cite{caloN} described by the Hamiltonian
\begin{equation}
H_{C}=\sum_{i=1}^{N}\frac{p_{i}^{2}}{2}+\frac{1}{2}\sum_{i\neq j}^{N}\frac{g%
}{(x_{i}-x_{j})^{2}},  \label{calham}
\end{equation}

\noindent with $g\in \mathbb{R}$ being the coupling constant. Marchioro \cite%
{marchi, marcal} investigated thereafter the classical analogue of these
models obtaining a solution to which we will appeal below. The integrability
of these classical counterparts was established later by Moser \cite{moser},
using a Lax pair consisting of matrices $L,M$, with entries
\begin{eqnarray}
L_{ij} &=&p_{i}\delta _{ij}+\frac{\imath \sqrt{g}}{x_{i}-x_{j}}(1-\delta
_{ij}),  \label{calax} \\
M_{ij} &=&\sum_{k\neq i}^{N}\frac{\imath \sqrt{g}}{(x_{i}-x_{k})^{2}}\delta
_{ij}-\frac{\imath \sqrt{g}}{(x_{i}-x_{j})^{2}}(1-\delta _{ij}),
\label{camax}
\end{eqnarray}

\noindent constructed in such a manner that the Lax equation
\begin{equation}
\frac{dL}{dt}+[M,L]=0
\end{equation}

\noindent becomes equivalent to the Calogero equations of motion,
\begin{equation}
\ddot{x}_{i}=\sum_{j\neq i}^{N}\frac{2g}{(x_{i}-x_{j})^{3}}.  \label{caleq}
\end{equation}

\noindent We use the notation $\imath \equiv \sqrt{-1}$ throughout the
manuscript and abbreviate time derivatives as usual\ by $dx_{i}/dt=\dot{x}%
_{i}$ and $d^{2}x_{i}/dt^{2}=\ddot{x}_{i}$. Integrability follows in the
standard fashion by noting that all quantities of the for $I_{n}=\limfunc{tr}%
(L^{n})$ $/n~$are integrals of motion and conserved in time by construction.

Calogero systems have become very important in theoretical physics, having
been explored in various contexts ranging from condensed matter physics to
cosmology, e.g. \cite{rmatrix, qhall, black}. The main focus of our interest
here are the complex extensions which have been studied recently in
connection with $\mathcal{PT}$-symmetric models \cite{basu, taterz,
calfring, jain,fringz}.

The idea of exploiting $\mathcal{PT}$-symmetry in order to obtain models
with real energies can be adapted to classical systems as well and has been
used to formulate various complex extensions of nonlinear wave equations,
such as the Korteweg-de Vries (KdV) and Burgers equations \cite{benderfein,
benderbcf, kdvfring, afint}. In the classical case the reality of the energy
is ensured in an even simpler way, as in that case the $\mathcal{PT}$%
-symmetry of the Hamiltonian is sufficient. Remarkably these systems allow
for the existence of solitons and compacton solutions \cite{bendercooper,
afcomp}.

Here we shall explore $\mathcal{PT}$-symmetry in a context where the complex
extensions or deformations do not need to be imposed artificially, but
instead we investigate whether this symmetry is already naturally present in
the system, albeit hidden. To achieve this goal we exploit the fact that
nonlinear equations, such as Benjamin-Ono and Boussinesq, can be associated
to Calogero particle systems. We explore these connections and are then
naturally led to complex $\mathcal{PT}$-symmetric Calogero systems.

In the next section we shall demonstrate how a complex one-dimensional $%
\mathcal{PT}$-symmetric Calogero system is embedded in a real solitonic
solution of the Benjamin-Ono wave equation and how constrained $\mathcal{PT}$%
-symmetric Calogero particles emerge from real solutions of the Boussinesq
equation. Thereafter we construct the explicit solution of the
three-particle configuration with the aforementioned constraint and show
that the resulting motion, unlike in the unconstrained situation, cannot be
restricted to the real line. We shall also establish that a subclass of this
constrained Calogero motion is related to the poles in the solution of
different nonlinear KdV-like differential equation. The relation of these
complex particles with previously obtained $\mathcal{PT}$-symmetric complex
extensions of the Calogero model \cite{afindia} is discussed in section 5,
where we demonstrate that they are different from those proposed here. Our
conclusions are drawn in section 6.

\section{Poles of nonlinear waves as interacting particles}

\label{waves}

The assumption of rational real valued functions as multi-soliton solutions
of nonlinear wave equations was studied more than three decades ago by
various authors, see e.g. \cite{amm}. We take some of these findings as a
setting for the problem at hand. In order to illustrate the key idea we
present what is probably the simplest scenario in which corpuscular objects
emerge as poles of nonlinear waves, namely in the Burgers equation
\begin{equation}
u_{t}+\alpha u_{xx}+\beta (u^{2})_{x}=0.  \label{burq}
\end{equation}
Assuming that this equation admits rational solutions of the form
\begin{equation}
u(x,t)=\frac{2\alpha }{\beta }\sum_{i=1}^{N}\frac{1}{x-x_{i}(t)},
\label{anburg}
\end{equation}

\noindent it is straightforward to see that surprisingly the $N$ poles
interact with each other through a Coulombic inverse square force
\begin{equation}
\ddot{x}_{i}(t)=-2\alpha \sum_{j\neq i}^{N}\frac{1}{[x_{i}(t)-x_{j}(t)]^{2}}.
\label{ecoul}
\end{equation}

This pole structure survives even after making modifications in the ansatz
for the wave equation, although the nature of the interaction may change. By
acting on the second derivative in Burgers equation with a Hilbert transform
\begin{equation}
\hat{H}u(x)=\frac{1}{\pi }PV\int_{-\infty }^{+\infty }dz\frac{u(z)}{z-x},
\end{equation}

\noindent we obtain the Benjamin-Ono equation \cite{benjamin,ono}
\begin{equation}
u_{t}+\alpha \hat{H}u_{xx}+\beta (u^{2})_{x}=0.  \label{boq}
\end{equation}

As shown in \cite{chenlp}, the ansatz proposed for the equation above which
will allow for similar conclusions has a slightly different form,
\begin{equation}
u(x,t)=\frac{\alpha }{\beta }\sum_{k=1}^{N}\left( \frac{\imath }{x-z_{k}(t)}-%
\frac{\imath }{x-z_{k}^{\ast }(t)}\right)  \label{anbo}
\end{equation}

\noindent being, however, still a real valued solution with the only
restriction that the complex poles satisfy complex Calogero equations of
motion
\begin{equation}
\ddot{z}_{k}(t)=8\alpha ^{2}\sum_{k\neq j}^{N}\frac{1}{%
(z_{k}(t)-z_{j}(t))^{3}}.  \label{eccal}
\end{equation}

Note that there is a difference in the power laws appearing in (\ref{ecoul})
and (\ref{eccal}), but more importantly that equation (\ref{anburg}) has
real poles, whereas (\ref{anbo}) has complex ones. We stress once more that
the field $u(x,t)$ is real in both cases. Hence, this viewpoint provides a
nontrivial mechanism which leads to particle systems defined in the complex
plane.

Interesting observations of this kind can be made for other nonlinear
equations as well, but not always will the ansatz work directly, that is
without any further requirements as in the previous cases. In some
situations additional conditions might be necessary. Examples of nonlinear
integrable wave equations for which such type of constraints occur are the
KdV and the Boussinesq equations,
\begin{equation}
u_{t}+\left( \alpha u_{xx}+\beta u^{2}\right) _{x}=0\;\;\;\;\;\;\;\text{and}%
\;\;\;\;\;\;\;u_{tt}+\left( \alpha u_{xx}+\beta u^{2}-\gamma u\right)
_{xx}=0,  \label{waveq}
\end{equation}

\noindent respectively. For both of these equations one can have
\textquotedblleft $N$-soliton\textquotedblright\ solutions\footnote{%
Soliton is to be understood here in a very loose sense in analogy to the
Painlev\'{e} type ideology of indistructable poles. In the strict sense not
all solution possess the $N$-soliton solution characteristic, that is moving
with a preserved shape and regaining it after scattering though each other.}
of the form
\begin{equation}
u(x,t)=-6\frac{\alpha }{\beta }\sum_{k=1}^{N}\frac{1}{(x-x_{k}(t))^{2}},
\label{ssol}
\end{equation}

\noindent as long as in each case two sets of constraints are satisfied
\begin{equation}
{\dot{x}}_{k}(t)=-12\alpha \sum_{j\neq
k}^{N}(x_{k}(t)-x_{j}(t))^{-2}\;\;,\;\;\;\;\;\;\;\;\;\;\;\;\;\;\;\;\;0=%
\sum_{j\neq k}^{N}(x_{k}(t)-x_{j}(t))^{-3},\;\;\;\;\;\;\;\;\;\;\;\;\;
\label{ckdv}
\end{equation}

\noindent and
\begin{equation}
{\ddot{x}}_{k}(t)=-24\alpha \sum_{j\neq
k}^{N}(x_{k}(t)-x_{j}(t))^{-3}\;\;,\;\;\;\;\;\;\;\;\;{\dot{x}}%
_{k}(t)^{2}=12\alpha \sum_{j\neq k}^{N}(x_{k}(t)-x_{j}(t))^{-2}+\gamma ,
\label{cbouss}
\end{equation}
respectively. Naturally these constraints might be incompatible or admit no
solution at all, in which case (\ref{ssol}) would of course not constitute a
solution for the wave equations (\ref{waveq}). Notice that if the $x_{k}(t)$
are real or come in complex conjugate pairs the solution (\ref{ssol}) for
the corresponding wave equations is still real.

Airault, McKean and Moser provided a general criterium, which allows us to
view these equations from an entirely different perspective, namely to
regard them as constrained multi-particle systems \cite{amm}:

\textit{Given a multi-particle Hamiltonian $H(x_{1},...,x_{N},\dot{x}%
_{1},...,\dot{x}_{N})$ with flow $x_{i}=\partial H/\partial \dot{x}_{i}$ and
$\dot{x}_{i}=-\partial H/\partial x_{i}$ together with conserved charges $%
I_{n}$ in involution with $H$, i.e.~vanishing Poisson brackets $%
\{H,I_{n}\}=0 $, then the locus of $grad(I_{n})=0$ is invariant with respect
to time evolution. Thus it is permitted to restrict the flow to that locus
provided it is not empty.}

Taking the Hamiltonian to be the Calogero Hamiltonian $H_{C}$ it is well
known that one may construct the corresponding conserved quantities from the
Calogero Lax operator (\ref{calax}) as mentioned from $I_{n}=\text{tr}%
(L^{n})/n$. The first of these charges is just the total momentum, the next
is the Hamltonian followed by non trivial ones
\begin{equation}
I_{1}=\sum_{i=1}^{N}p_{i},\qquad I_{2}=H_{C}(g),\qquad I_{3}=\frac{1}{3}%
\sum_{i=1}^{N}p_{i}^{3}+g\sum_{i\neq j}^{N}\frac{p_{i}+p_{j}}{%
(x_{i}-x_{j})^{2}}\;,\ldots   \label{F123}
\end{equation}

According to the above mentioned criterium we may therefore consider an $%
I_{3}$-flow restricted to the locus defined by $\limfunc{grad}(I_{2})=0$ or
an $I_{2}$-flow subject to the constraint $\limfunc{grad}(I_{3}-\gamma
I_{1})=0$. Remarkably it turns out that the former viewpoint corresponds
exactly to the set of equations (\ref{ckdv}), whereas the latter to (\ref%
{cbouss}) when we identify the coupling constant as $g=-12\alpha $. Thus the
solutions of the Boussinesq equation are related to the constrained Calogero
Hamiltonian flow, whereas the KdV soliton solutions arise from an $I_{3}$%
-flow subject to constraining equations derived from the Calogero
Hamiltonian.

As our main focus is on the Calogero Hamiltonian flow and its possible
complexifications we shall concentrate on possible solutions of the systems (%
\ref{cbouss}) and investigate whether these type of equations allow for
nontrivial solutions or whether they are empty. It will be instructive to
commence by looking first at the unconstrained system. The classical
solutions of a two-particle Calogero problem are given by
\begin{equation}
x_{1,2}(t)=2R(t)\pm \sqrt{\frac{g}{E}+4E(t-t_{0})^{2}},  \label{2un}
\end{equation}

\noindent with $E,t_{0}$ being initial conditions and $\dot{R}(t)=0$ the
centre of mass velocity. Relaxing this condition by allowing boosts will
only shifts the energy scale since the total momentum is conserved.
Depending therefore on the initial conditions we may have either real or
complex solutions.

The three particle model, i.e.~taking $N=3$ in (\ref{calham}), is slightly
more complicated. Marchioro \cite{marchi} found the general solution by
expressing the dynamical variables in terms of Jacobi relative coordinates $%
R $, $X$, $Y$ in polar form via the transformations $%
R(t)=(x_{1}(t)+x_{2}(t)+x_{3}(t))/3$, $X(t)=r(t)\sin \phi
(t)=(x_{1}(t)-x_{2}(t))/\sqrt{2}$ and $Y(t)=r(t)\cos \phi
(t)=(x_{1}(t)+x_{2}(t)-2x_{3}(t))/\sqrt{6}$. The variables may then be
separated and the resulting equations are solved by
\begin{eqnarray}
x_{1,2}(t) &=&R(t)+\frac{1}{\sqrt{6}}r(t)\cos {\phi (t)}\pm \frac{1}{\sqrt{2}%
}r(t)\sin {\phi (t)},  \label{3uncsol1} \\
x_{3}(t) &=&R(t)-\frac{2}{\sqrt{6}}r(t)\cos {\phi (t)},  \label{3uncsol2}
\end{eqnarray}

\noindent where
\begin{eqnarray}
R(t) &=&R_{0}+t\tilde{R}_{0}  \label{218} \\
r(t) &=&\sqrt{\frac{B^{2}}{E}+2E(t-t_{0})^{2}}, \\
\phi (t) &=&\frac{1}{3}\cos ^{-1}{\left\{ \varphi _{0}\sin {\left[ \sin ^{-1}%
{\left( \varphi _{0}\cos {3\phi _{0}}\right) -3\tan ^{-1}{\left( \frac{\sqrt{%
2}E}{B}(t-t_{0})\right) }}\right] }\right\} }.
\end{eqnarray}

\noindent The solutions involve $7$ free parameters: The total energy $E$,
the angular momentum type constant of motion $B$, the integration constants $%
t_{0}$, $\phi _{0}$, $R_{0}$, $\tilde{R}_{0}$ and the coupling constant $g$,
with the abbreviation $\varphi _{0}=\sqrt{1-9g/2B^{2}}$. We note that,
depending on the choice of these parameters, both real and complex solutions
are admissible, a feature which might not hold for the Calogero system
restricted to an invariant submanifold.

Let us now elaborate further on the connection between the field equations
and the particle system and restrict the general solution (\ref{3uncsol1})-(%
\ref{218}) by switching on the additional constraints in (\ref{cbouss}) and
subsequently study the effect on the soliton solutions of the nonlinear wave
equation. Notice that the second constraint in (\ref{cbouss}) can be viewed
as setting the difference between the kinetic and potential energy of each
particle to a constant. Adding all of these equations we obtain $%
H_{C}=N\gamma /2$, which provides a direct interpretation of the constant $%
\gamma $ in the Boussinesq equation as being proportional to the total
energy of the Calogero model.

\section{The motion of Boussinesq singularities}

\label{boussing}

The two particle system, i.e.~$N=2$, is evidently the simplest $I_{2}$%
-Calogero flow constrained with $\limfunc{grad}(I_{3}-\gamma I_{1})=0$ as
specified in (\ref{cbouss}). The solution for this system was already
provided in \cite{amm},
\begin{equation}
x_{1,2}(t)=\kappa \pm \sqrt{\gamma (t-\tilde{\kappa})^{2}-3\alpha /\gamma },
\label{two}
\end{equation}

\noindent with $\kappa ,\tilde{\kappa}$ taken to be real constants. In fact
this solution is not very different from the unconstrained motion shown in
the previous section (\ref{2un}). The restricted one may be obtained via an
identification between the coupling constant and the parameter in the
Boussinesq equation as $\kappa =2R(t)$, $E=\gamma /4$, $\tilde{\kappa}=t_{0}$
and $g=-3\alpha /4$. The two soliton solution for the Boussinesq equation (%
\ref{ssol}) then acquires the form
\begin{equation}
u(x,t)=-12\frac{\alpha }{\beta }\gamma \frac{\gamma (x-\kappa )^{2}+\gamma
^{2}(t-\tilde{\kappa})^{2}-3\alpha }{[\gamma (x-\kappa )^{2}-\gamma ^{2}(t-%
\tilde{\kappa})^{2}+3\alpha ]^{2}},  \label{11}
\end{equation}

\noindent which, in principle, is still real-valued when keeping the
constants to be real. When inspecting (\ref{11}) it is easy to see that the
two singularities repel each other on the $x$-axis as time evolves, thus
mimicking a repulsive scattering process. However, we may change the overall
behaviour substantially when we allow the integration constants to be
complex, such that the singularities become regularized. In that case we
observe a typical solitonic scattering behaviour, i.e.~two wave packets
keeping their overall shape while evolving in time and when passing though
each other regaining their shape when the scattering process is finished,
albeit with complex amplitude. A special type of complexification occurs
when we take the integration constants $\kappa ,\tilde{\kappa}$ to be purely
imaginary, in which case (\ref{11}) becomes a solution for the $\mathcal{PT}$%
-symmetrically constrained Boussinesq equation, with $\mathcal{PT}%
:x\rightarrow -x,t\rightarrow -t,u\rightarrow u$. We depict the described
behaviour in figure 1 for some special choices of the parameters.

\vspace{-0.8cm}

\begin{center}
\includegraphics[width=16cm]{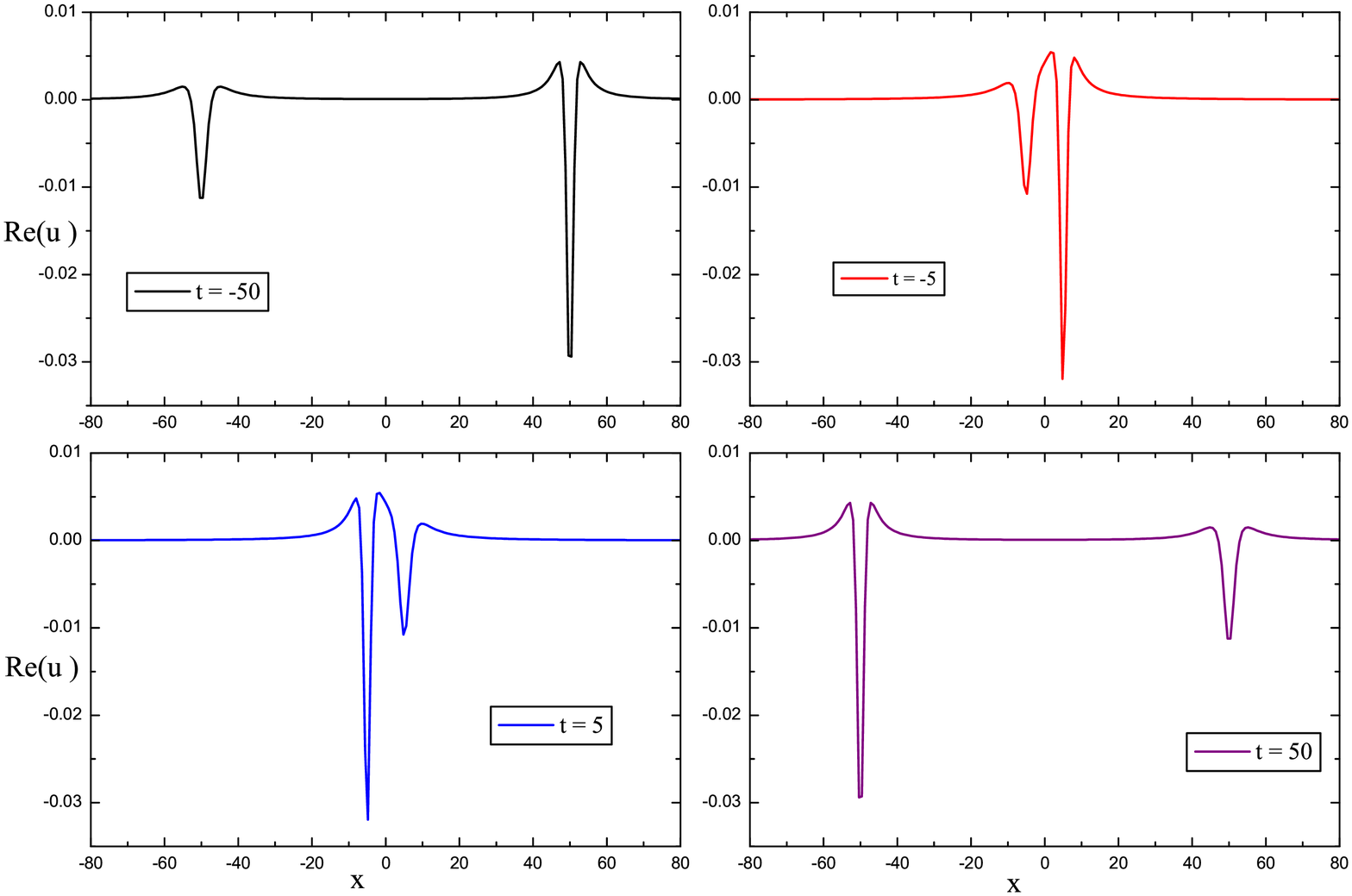}
\end{center}

\vspace{-0.8cm}

\noindent {\textbf{Figure 1:} {\small {\ Time evolution of the real part of
the constraint Boussinesq two soliton solution (\ref{11}) with $\kappa
=\imath 2.3$ , $\tilde{\kappa}=-\imath 0.6$, $\alpha =-1/6,${\small {$\beta
=5/8$}} and }}}$\gamma ${{\small {{\small {$=1$}}. }} }

For larger numbers of particles the solutions have not been investigated and
it is not even clear whether the locus of interest is empty or not. Let us
therefore embark on solving this problem systematically. Unfortunately we
can not simply imitate Marchioro's method of separating variables as the
additional constraints will destroy this possibility. However, we notice
that (\ref{cbouss}) can be represented in a different way more suited for
our purposes. Differentiating the second set of equations in (\ref{cbouss})
and making use of the first one, we arrive at the set of expressions
\begin{equation}
\sum_{k\neq j}^{N}\frac{(\dot{x}_{k}(t)+\dot{x}_{j}(t))}{%
(x_{k}(t)-x_{j}(t))^{3}}=0,  \label{cboussalt}
\end{equation}

\noindent which are therefore consistency equations of the other two.

We now focus on the case $N=3$. Inspired by the general solution of the
unconstrained three particle solution (\ref{3uncsol1}) and (\ref{3uncsol2}),
we adopt an ansatz of the general form
\begin{eqnarray}
x_{1,2}(t) &=&A_{0}(t)+A_{1}(t)\pm A_{2}(t),  \label{anx1} \\
x_{3}(t) &=&A_{0}(t)+\lambda A_{1}(t),  \label{anx2}
\end{eqnarray}

\noindent with $A_{i}(t)$, $i=0,1,2$ being some unknown functions and $%
\lambda $ a free constant parameter. We note that $\lambda \neq 1$, since
otherwise the three coordinates could be expressed in terms of only two
linearly independent functions, $A_{0}(t)+A_{1}(t)$ and $A_{2}(t)$, and we
would not able to express the normal mode like functions $A_{i}(t)$ in terms
of the original coordinates $x_{i}(t)$. Calogero's choice, $\lambda =-2$, in
equation (\ref{3uncsol2}), allows an elegant map of Cartesian coordinates
into Jacobi's relative coordinates, but other possibilities might be more
convenient in the present situation. Here we keep $\lambda $ to be free for
the time being.

Substituting this ansatz for the $x_{i}(t)$ into the second set of equations
in (\ref{cbouss}) and using the compatibility equation (\ref{cboussalt}), we
are led to six coupled first order differential equations for the unknown
functions $A_{0}(t),A_{1}(t),A_{2}(t)$%
\begin{eqnarray}
\frac{(\dot{A}_{0}(t)+\lambda \dot{A}_{1}(t))^{2}-\gamma }{2g}+\frac{1}{%
2A_{+}(t)^{2}}+\frac{1}{2A_{-}(t)^{2}} &=&0,  \label{da3} \\
\frac{(\dot{A}_{0}(t)+\dot{A}_{1}(t)\pm \dot{A}_{2}(t))^{2}-\gamma }{2g}+%
\frac{1}{8A_{2}(t)^{2}}+\frac{1}{2A_{\mp }(t)^{2}} &=&0,  \label{da4} \\
\frac{2\dot{A}_{0}(t)+(\lambda +1)\dot{A}_{1}(t)+\dot{A}_{2}(t)}{A_{-}(t)^{3}%
}-\frac{2\dot{A}_{0}(t)+(\lambda +1)\dot{A}_{1}(t)-\dot{A}_{2}(t)}{%
A_{+}(t)^{3}} &=&0,  \label{da1} \\
\frac{\dot{A}_{0}(t)+\dot{A}_{1}(t)}{4A_{2}(t)^{3}}+\frac{2\dot{A}%
_{0}(t)+(\lambda +1)\dot{A}_{1}(t)\pm \dot{A}_{2}(t)}{A_{\mp }(t)^{3}} &=&0.
\label{da2}
\end{eqnarray}

\noindent For convenience we made the identifications $A_{\pm
}(t)=A_{2}(t)\pm (\lambda -1)A_{1}(t)$.

From the latter set of equations above, (\ref{da1}) and (\ref{da2}), we can
now eliminate two of the first derivatives together with the use of the
conservation of momentum. Depending on the choice, the remaining $\dot{A}%
_{i}(t)$ are eliminated with the help of the first three equations (\ref{da3}%
) and (\ref{da4}). The two equations left then become multiples of each
other depending only on $A_{1}(t)$ and $A_{2}(t)$. Subsequently we can
express $A_{2}(t)$, and consequently $\dot{A}_{0}(t),\dot{A}_{1}(t)$, in
terms of $A_{1}(t)$ as the only unknown quantity. In this manner we arrive
at
\begin{eqnarray}
A_{2}(t) &=&\frac{\sqrt{-g-4\gamma (\lambda -1)^{2}A_{1}(t)^{2}}}{2\sqrt{%
3\gamma }},  \label{a2} \\
\dot{A}_{0}(t) &=&\sqrt{\gamma }+\frac{3g\sqrt{\gamma }(2+\lambda )}{%
(\lambda -1)[g+16\gamma (\lambda -1)^{2}A_{1}(t)^{2}]},  \label{a0} \\
\dot{A}_{1}(t) &=&\frac{9g\sqrt{\gamma }}{(1-\lambda )[g+16\gamma (\lambda
-1)^{2}A_{1}(t)^{2}]},  \label{a1}
\end{eqnarray}%
with $g=-12\alpha $. This means that once we have solved the differential
equation (\ref{a1}) for $A_{1}(t)$ the complete solution is determined up to
the integration of $\dot{A}_{0}(t)$ in (\ref{a0}) and a simple substitution
in (\ref{a2}). In other words we have reduced the problem to solve the set
of coupled nonlinear equations (\ref{cbouss}) to solving one first order
nonlinear equation.

Let us now make a comment on the number of free parameters, that is
integration constants, occurring in this solution. In the original
formulation of the problem we have started with 3 second order differential
equations, so that we expect to have 6 integration constants for the
determination of $x_{1},x_{2}$ and $x_{3}$. However, together with the
additional 3 constraining equations this number is reduced to 3 free
parameters. Finally we can invoke the conservation of total momentum from (%
\ref{F123}), which yields $3\ddot{A}_{0}(t)+(\lambda +2)\ddot{A}_{1}(t)=0$
and we are left with only 2 free parameters. We choose them here to be the
two arbitrary constants attributed to the integration of $\dot{A}_{0}(t)$ in
(\ref{a0}) and $\dot{A}_{1}(t)$ in (\ref{a1}), respectively.

In turn this also means that, without loss of generality, we may freely
choose the constant $\lambda $ introduced in (\ref{anx2}). Indeed, keeping
it generic we observe that the solutions for the $A_{i}(t)$ do not depend on
it despite its explicit presence in the equations (\ref{a2}), (\ref{a0}) and
(\ref{a1}). The most convenient choice is to take $\lambda =-2$ as in that
case the equations simplify considerably.

Let us now solve (\ref{a2}), (\ref{a0}) and (\ref{a1}) and substitute the
result into the original expressions (\ref{anx1}) and (\ref{anx2}) in order
to see how the particles behave. We find
\begin{eqnarray}
x_{1,2}(t) &=&c_{0}+\sqrt{\gamma }t+\frac{1}{12}\left( \frac{g}{\xi (t)}-%
\frac{\xi (t)}{\gamma }\right) \pm \frac{\imath }{4\sqrt{3}}\left( \frac{g}{%
\xi (t)}+\frac{\xi (t)}{\gamma }\right) ,  \label{x12} \\
x_{3}(t) &=&c_{0}+\sqrt{\gamma }t-\frac{1}{6}\left( \frac{g}{\xi (t)}-\frac{%
\xi (t)}{\gamma }\right) ,  \label{x3}
\end{eqnarray}

\noindent where for convenience we introduced the abbreviation
\begin{equation}
\xi (t)=\left[ -54\gamma ^{2}(\sqrt{\gamma }gt+c_{1})+\sqrt{g^{3}\gamma
^{3}+[54\gamma ^{2}(\sqrt{\gamma }gt+c_{1})]^{2}}\right] ^{\frac{1}{3}}.
\label{xi}
\end{equation}

\noindent The above mentioned two freely choosable constants of integration
are denoted by $c_{0}$ and $c_{1}$. \ As in the two particle case, we may
once again compare this solution with the unconstrained one in (\ref%
{3uncsol1}), (\ref{3uncsol2}) when considering the Jacobi relative
coordinates
\begin{equation}
R(t)=c_{0}+t\sqrt{\gamma },\quad r^{2}(t)=-\frac{g}{6\gamma }\quad \text{%
and\quad }\tan {\phi (t)=i}\frac{{g\gamma +\xi }^{2}{(t)}}{{g\gamma -\xi }%
^{2}{(t)}}.  \label{Jacko}
\end{equation}%
We observe that the solution is now constrained to a circle in the $XY$%
-plane with real radius when $g\gamma \in \mathbb{R}^{-}$. The values for ${%
\phi (t)}$ lead to the most dramatic consequence, namely that the particles
are now forced to move in the complex plane, unlike as in unconstrained
Calogero system or the $N=2$ case where all options are open.

Interestingly, despite the poles being complex, we may still have real wave
solutions for the Boussinesq equation. Provided that $\xi (t),\gamma
,g,c_{0},c_{1}\in \mathbb{R}$ the pole $x_{3}(t)$ is obviously real whereas $%
x_{1}(t)$ and $x_{2}(t)$ are complex conjugate to each other, such that the
ansatz (\ref{ssol}) yields a real solution
\begin{eqnarray}
u(x,t) &=&-\frac{6\alpha }{\beta }\frac{1}{\left( \varphi -\frac{1}{6}\left(
\frac{g}{\xi (t)}-\frac{\xi (t)}{\gamma }\right) \right) ^{2}}+ \\
&+&\frac{216\alpha }{\beta }\gamma ^{2}\xi (t)^{2}\left[ \frac{g^{2}\gamma
^{2}-12g\gamma ^{2}\varphi \xi (t)-4\gamma (18\gamma \varphi ^{2}-g)\xi
(t)^{2}+12\gamma \varphi \xi (t)^{3}+\xi (t)^{4}}{(g^{2}\gamma ^{2}+6g\gamma
^{2}\varphi \xi (t)+\gamma (36\gamma \varphi ^{2}+g)\xi (t)^{2}-6\gamma
\varphi \xi (t)^{3}+\xi (t)^{4})^{2}}\right]  \notag
\end{eqnarray}

\noindent with $\varphi \equiv c_{0}+\sqrt{\gamma }t-x$.

Due to the non-meromorphic form of $\xi (t)$ it is not straightforward to
determine how the solutions transforms under a $\mathcal{PT}$%
-transformation. Nonetheless, the symmetry of the relevant combinations
appearing in (\ref{x12}) and (\ref{x3}) can be analyzed well for $%
c_{0},c_{1}\in i\mathbb{R}$ and $\gamma >0$\thinspace . In that case the
time reversal acts as $\mathcal{T}:\left( \frac{g}{\xi (t)}\pm \frac{\xi (t)%
}{\gamma }\right) \rightarrow \pm \left( \frac{g}{\xi (t)}\pm \frac{\xi (t)}{%
\gamma }\right) $, which implies $\mathcal{PT}:x_{i}(t)\rightarrow -x_{i}(t)$
for $i=1,2,3$. Thus, the solutions to the constrained problem are not only
complex, but in addition they can also be $\mathcal{PT}$-symmetric for
certain choices of the constants involved.

\section{Different types of constraints in nonlinear wave equations \label%
{calpart}}

It is clear from the above that the class of complex ($\mathcal{PT}$%
-symmetric) multi-particle systems which might arise from nonlinear wave
equations could be much larger. We shall demonstrate this by investigating
one further simple example which was previously studied in \cite{bowtell}
and also refer to the literature \cite{tzi} for additional examples. One
very easy nonlinear wave equations which, because of its simplicity, serves
as a very instructive toy model is
\begin{equation}
u_{t}+u_{x}+u^{2}=0.  \label{bowq}
\end{equation}
We may now proceed as above and seek for a suitable ansatz to solve this
equation, possibly leading to some constraining equations in form a
multi-particle systems. Making therefore a similar ansatz for $u(x,t)$ as in
(\ref{anbo}) or (\ref{ssol}) we take
\begin{equation}
u(x,t)=\sum_{i=1}^{N}\frac{1-\dot{z}_{i}(t)}{x-z_{i}(t)}.  \label{an1}
\end{equation}
It is then easy to verify that this solves the nonlinear equation (\ref{bowq}%
) provided the $z_{i}(t)$ obey the constraints
\begin{equation}
\ddot{z}_{i}(t)=2\sum_{j\neq i}^{N}\frac{(1-\dot{z}_{i}(t))(1-\dot{z}_{j}(t))%
}{z_{i}(t)-z_{j}(t)}.  \label{const}
\end{equation}

\noindent We could now proceed as in the previous section and try to solve
this differential equation, but in this case we may appeal to the general
solution already provided in \cite{bowtell}, where it was found that
\begin{equation}
u(x,t)=\frac{f(x-t)}{1+tf(x-t)}.  \label{an2}
\end{equation}
solves (\ref{bowq}) for any arbitrary function $f(x)$ with initial condition
$u(x,0)=f(x)$. Comparing (\ref{an2}) and (\ref{an1}) it is clear that the $%
z_{i}(t)$ can be interpreted as the poles in (\ref{an2}), which becomes
singular when $x\rightarrow z_{i}(t)=t+f_{i}^{-1}(-1/t)$, with $i\in \{1,N\}$
labeling the different branches which could result when assuming that $f$ is
invertible but not necessarily injectively. Making now the concrete choice
for $f$ to be rational of the form
\begin{equation}
f(x)=\sum\limits_{i=1}^{N}\frac{a_{i}}{\alpha _{i}-x},\qquad \qquad \text{%
with }\alpha _{i},a_{i}\in \mathbb{C},  \label{rat}
\end{equation}

\noindent we can determine the poles concretely by inverting this function.
First of all we obtain from the initial condition that
\begin{equation}
z_{i}(0)=\alpha _{i},\qquad \dot{z}_{i}(0)=1+a_{i}\qquad \text{and\qquad }%
\ddot{z}_{i}(0)=\sum_{j\neq i}^{N}\frac{2a_{i}a_{j}}{\alpha _{i}-\alpha _{j}}%
.
\end{equation}
The first two conditions simply follow from the comparison of (\ref{an2})
and (\ref{an1}), but also follow, as so does the latter, from taking the
appropriate limit in (\ref{rat}). We note that the total momentum is
conserved for this system $\sum_{i=1}^{N}\dot{z}_{i}(t)=N+%
\sum_{i=1}^{N}a_{i} $.

Let us now see how to obtain explicit expressions for the poles. Inverting (%
\ref{rat}) for $N=2$ it is easy to find that for generic values of $t$ the
poles take on the form
\begin{equation}
z_{1,2}(t)=t+\frac{\bar{\alpha}_{12}}{2}+\frac{\bar{a}_{12}}{2}t\pm \frac{1}{%
2}\sqrt{\alpha _{12}^{2}+2\alpha _{12}a_{12}t+\bar{a}_{12}^{2}t^{2}},
\end{equation}
where we introduced the notation $\alpha _{ij}=\alpha _{i}-\alpha _{j}$, $%
\bar{\alpha}_{ij}=\alpha _{i}+\alpha _{j}$ and analogously for $\alpha
\rightarrow a$. We note that in the case $N=2$ the constraint (\ref{const})
can be changed into two-particle Calogero systems constraint with the
identification $g=a_{1}a_{2}\alpha _{12}^{2}$.

Next we consider the case $N=3$ for which we obtain the solution
\begin{eqnarray}
z_{1}(t) &=&t-\frac{a(t)}{3}+s_{+}(t)+s_{-}(t)  \label{3bow} \\
z_{2,3}(t) &=&t-\frac{a(t)}{3}-\frac{1}{2}\left[ s_{+}(t)+s_{-}(t)\right]
\pm \imath \frac{\sqrt{3}}{2}\left[ s_{+}(t)-s_{-}(t)\right] ,  \label{3bow3}
\end{eqnarray}
where we abbreviated
\begin{eqnarray}
s_{\pm }(t) &=&\left[ r(t)\pm \sqrt{r^{2}(t)+q^{3}(t)}\right] ^{1/3},\quad \\
r(t) &=&\frac{9a(t)b(t)-27c(t)-2a^{3}(t)}{54},\quad q(t)=\frac{3b(t)-a^{2}(t)%
}{9},\quad \\
a(t) &=&-a_{1}-\alpha _{2}-\alpha _{3}-t(a_{1}+a_{2}+a_{3}), \\
b(t) &=&\alpha _{1}\alpha _{2}+\alpha _{2}\alpha _{3}+\alpha _{1}\alpha
_{3}+t[a_{1}\bar{\alpha}_{23}+a_{2}\bar{\alpha}_{31}+a_{3}\bar{\alpha}%
_{21}],\quad \\
c(t) &=&-t(a_{1}\alpha _{2}\alpha _{3}+a_{2}\alpha _{3}\alpha
_{1}+a_{3}\alpha _{1}\alpha _{2})-\alpha _{1}\alpha _{2}\alpha _{3}.
\end{eqnarray}
In terms of Jacobi's relative coordinates this becomes
\begin{equation}
R(t)=t-\frac{1}{3}a(t),\quad r^{2}(t)=6s_{+}(t)s_{-}(t)\quad \text{and\quad }%
\tan {\phi (t)=i}\frac{s_{-}(t)-s_{+}(t)}{s_{-}(t)+s_{+}(t)},
\end{equation}
which makes a direct comparison with the constrained Calogero system (\ref%
{Jacko}) straightforward. As the system (\ref{3bow}), (\ref{3bow3}) involves
more free parameters than the constrained Calogero system (\ref{Jacko}), we
expect to observe some relations between the parameters $\alpha _{i},a_{i}$
to produce the right number of free parameters. Indeed, we find that for
\begin{equation}
a_{i}=-\frac{g}{2}\prod\limits_{j\neq i}(\alpha _{i}-\alpha _{j})^{-2}
\end{equation}
and the additional constraints
\begin{equation}
c_{0}=\frac{1}{3}\sum\limits_{i=1}^{3}\alpha _{i},\quad c_{1}=\frac{2}{27}%
\dprod\limits_{\substack{ 1\leq j<k\leq 3  \\ j,k\neq l}}(\alpha _{j}+\alpha
_{k}-2\alpha _{l}),\quad g=4\sum\limits_{\substack{ i=1  \\ i<j}}^{3}\alpha
_{i}\alpha _{j}-\alpha _{i}^{2},\quad \gamma =1.
\end{equation}
the two systems become identical. Thus we have obtained an identical
singularity structure for two quite different nonlinear wave equations.

\section{ Complex Calogero models and $\mathcal{PT}$-deformations \label%
{compcal}}

We have demonstrated in section \ref{boussing} that the solutions of the
constrained Calogero models are intrinsically of a complex nature. As there
have been various proposals before in the literature suggesting complex
Calogero systems in form of $\mathcal{PT}$-symmetric deformations, we shall
now compare them with the above outcome. We will argue that the deformations
presented here are new and different to those suggested up to now.

The simplest $\mathcal{PT}$-deformation of any model is obtained just by
adding a $\mathcal{PT}$-invariant term to the original Hamiltonian. For a
many-body situation, this was proposed for the first time in the framework
of $A_{n}$ Calogero models by introducing the Hamiltonian \cite{basu}
\begin{equation}
H(q,p)=H_{C}(q,p)+\sum_{i\neq j}^{N}\frac{\imath \tilde{g}p_{i}}{%
(x_{i}-x_{j})^{2}}.  \label{basuham}
\end{equation}

\noindent In \cite{calfring} it was shown that this simply corresponds to
shifting the momenta in the standard Calogero Hamiltonian together with a
re-definition of the coupling constant, which means the above construction
is certainly quite different from the proposal (\ref{basuham}).

The second type of deformation \cite{fringz} consists of replacing directly
the set of $\ell $-dynamical variables $q=\{q_{1},\ldots ,q_{\ell }\}$ and
their conjugate momenta $p=\{p_{1},\ldots ,p_{\ell }\}$ by means of a
deformation map $\varepsilon :(q,p)\rightarrow (\tilde{q},\tilde{p})$,
whereby the map is constructed in such a way that the original invariance
under the Weyl group $\mathcal{W}$ is replaced by an invariance under a $%
\mathcal{PT}$-symmetrically deformed version of the Weyl group $\mathcal{%
W^{PT}}$. In terms of roots the map is defined by replacing each root $%
\alpha $ by a deformed counterpart $\tilde{\alpha}$ as $\varepsilon :\alpha
\rightarrow \tilde{\alpha}$, whereby the precise form of the deformation
ensures the invariance under $\mathcal{W^{PT}}$ as specified in \cite{fringz}%
. Expanding the momenta as $p=\sum\nolimits_{i}\kappa _{i}\alpha _{i}$, with
$\kappa _{i}\in \mathbb{R}$, this means for the Calogero Hamiltonian
\begin{equation}
\varepsilon :H_{C}(q,p)\rightarrow H_{\mathcal{PT}}(\tilde{q},\tilde{p})=%
\frac{1}{2}\sum\limits_{i,j}\kappa _{i}\kappa _{j}\tilde{\alpha}_{i}\tilde{%
\alpha}_{j}+\frac{1}{2}\sum_{\tilde{\alpha}\in \tilde{\Delta}}\frac{g}{(%
\tilde{\alpha}\cdot q)^{2}}.
\end{equation}
In order to find the concrete forms for $\tilde{q}$ and $\tilde{p}$ we need
to be more specific about the algebras involved. Let us therefore examine
the models based on the rank $2$ algebras $A_{2}$, $B_{2}$ and $G_{2}$.
Depending on the dimensionality of the representation for the simple roots,
we obtain either a two or a three particle systems and may therefore compare
with the solutions found in the previous sections. In all cases the
deformations of the simple roots $\alpha _{1}$ and $\alpha _{2}$ take on the
general form
\begin{equation}
\quad \tilde{\alpha}_{1}(\varepsilon )=R(\varepsilon )\alpha
_{1}+iI(\varepsilon )K_{12}\lambda _{2},\qquad \text{and\qquad }\tilde{\alpha%
}_{2}(\varepsilon )=R(\varepsilon )\alpha _{2}-iI(\varepsilon )K_{21}\lambda
_{1},  \label{defa}
\end{equation}
with $\lambda _{1}$, $\lambda _{2}$ being fundamental weights obeying $%
2\lambda _{i}\cdot \alpha _{j}/\alpha _{j}^{2}=\delta _{ij}$, the functions $%
R(\varepsilon )$, $I(\varepsilon )$ satisfy $\lim_{\varepsilon \rightarrow
0}R(\epsilon )=1$, $\lim_{\varepsilon \rightarrow 0}I(\epsilon )=0$ and $%
K_{ij}=2\alpha _{i}\cdot \alpha _{j}/\alpha _{j}^{2}$ are the entries of the
Cartan matrix. Let us now take the following two dimensional representations
for the simple roots and fundamental weights
\begin{equation}
\begin{array}{lllll}
A_{2}: & \alpha _{1}=(1,-\sqrt{3}),~~ & \alpha _{2}=(1,\sqrt{3}),~~ &
\lambda _{1}=\frac{2}{3}\alpha _{1}+\frac{1}{3}\alpha _{2},~~ & \lambda _{2}=%
\frac{1}{3}\alpha _{1}+\frac{2}{3}\alpha _{2}, \\
B_{2}: & \alpha _{1}=(1,-1),~~ & \alpha _{2}=(0,1),~~ & \lambda _{1}=\alpha
_{1}+\alpha _{2},~~ & \lambda _{2}=\frac{1}{2}\alpha _{1}+\alpha _{2}, \\
G_{2}: & \alpha _{1}=(-\sqrt{\frac{3}{2}},\sqrt{\frac{1}{2}}),~~ & \alpha
_{2}=(\sqrt{6},0),~~ & \lambda _{1}=3\alpha _{1}+\alpha _{2},~~ & \lambda
_{2}=3\alpha _{1}+2\alpha _{2}.%
\end{array}
\label{rep2}
\end{equation}

We easily verify that this reproduces the correct entries for the Cartan
matrices $A_{2}:K_{11}=K_{22}=2$, $K_{12}=K_{21}=-1$, $B_{2}:K_{11}=K_{22}=2$%
, $K_{12}/2=K_{21}=-1$ and $G_{2}:K_{11}=K_{22}=2$, $K_{12}=K_{21}/3=-1$.
Having constructed the deformed roots we compute next the deformed conjugate
momenta and coordinates. In the representations (\ref{rep2}) the kinetic
energy term changes just by an overall factor as
\begin{equation}
\tilde{p}^{2}=\left[ R(\varepsilon )-\nu _{\mathbf{g}}^{2}I(\varepsilon )%
\right] p^{2}\qquad \text{with }\nu _{A_{2}}=1/\sqrt{3},\nu _{B_{2}}=1,\nu
_{G_{2}}=-\sqrt{3}.  \label{ppt}
\end{equation}%
The specific choice $R(\varepsilon )=\cosh \varepsilon $ and $I(\varepsilon
)=\nu _{\mathbf{g}}^{-2}\sinh \varepsilon $, used in \cite{fringz}, keeps
the kinetic energy term completely invariant, in the sense that the original
and deformed momenta are identical. The dual canonical coordinates $\tilde{q}
$ are computed from
\begin{equation}
\tilde{\alpha}\cdot q=\tilde{q}\cdot \alpha ,\qquad \alpha ,q\in \mathbb{R}%
\text{, }\tilde{\alpha},\tilde{q}\in \mathbb{R}\oplus i\mathbb{R}.
\end{equation}%
We find
\begin{equation}
\quad \tilde{q}_{1}=R(\varepsilon )q_{1}+i\nu _{\mathbf{g}}I(\varepsilon
)q_{2},\qquad \text{and\qquad }\tilde{q}_{2}=R(\varepsilon )q_{2}-i\nu _{%
\mathbf{g}}I(\varepsilon )q_{1}.  \label{qq}
\end{equation}%
We will now argue that (\ref{qq}) is always different from the constrained
two particle solution of the Calogero model (\ref{two}). In order to see
this we recall first of all that for the solution to be $\mathcal{PT}$%
-symmetric we require $\kappa ,\tilde{\kappa}\in i\mathbb{R}.$ Equating now
the sums $x_{1}+x_{2}=\tilde{q}_{1}+\tilde{q}_{2}$ we conclude that $%
q_{1}(t)=-q_{2}(t)=-\kappa /\nu _{\mathbf{g}}I(\varepsilon )=\limfunc{const}$%
. Next we compute $(x_{1}-x_{2})^{2}$, which yields
\begin{equation}
\gamma (t-\tilde{\kappa})^{2}-3\alpha /\gamma =2R^{2}(\varepsilon
)q_{1}^{2}(t).
\end{equation}%
This equation is inconsistent as the right had side is real and time
independent, a condition which can not be achieved for the left hand side.
This proves our statement that the deformation method suggested here is
genuinely different from the proposal in \cite{fringz} in the two particle
case.

Keeping the deformed roots to be of the form (\ref{defa}), the three
dimensional repesentations for the simple roots
\begin{equation}
A_{2}:\alpha _{1}=(1,-1,0),~~\alpha _{2}=(0,1,-1),~\quad G_{2}:\alpha
_{1}=(1,-1,0),~~\alpha _{2}=(-2,1,1),
\end{equation}%
yield the same result for the kinetic energy term (\ref{ppt}), but obviously
have to produce different dual canonical coordinates $\tilde{q}$. In this
case we obtain
\begin{eqnarray}
\quad \tilde{q}_{1} &=&R(\varepsilon )q_{1}+i\zeta _{\mathbf{g}%
}I(\varepsilon )(q_{2}-q_{3}),\qquad  \\
\tilde{q}_{2} &=&R(\varepsilon )q_{2}+i\zeta _{\mathbf{g}}I(\varepsilon
)(q_{3}-q_{1}), \\
\tilde{q}_{3} &=&R(\varepsilon )q_{3}+i\zeta _{\mathbf{g}}I(\varepsilon
)(q_{1}-q_{2}),
\end{eqnarray}%
where $\zeta _{A_{2}}=1/3$ and $\zeta _{G_{2}}=-1$. Equating these solutions
with (\ref{x12}), (\ref{x3}) and solving the resulting equations for the $%
q_{i}$ with $i=1,2,3$, it is easy to argue that the $q_{i}$ can not be made
real, which establishes the claim that the solutions are also intrisically
different for the three particle model.

\section{Conclusions}

\label{conc}

Hitherto there have been two different types of procedures to complexify
Calogero models. As explained in section 5 one may either add $\mathcal{PT}$%
-symmetric terms to the original Hamiltonian \cite{basu}, which have turned
out to be simple shifts in the momenta \cite{calfring} or one may directly
deform the root system on which the formulation of the model is based \cite%
{fringz}. In all these approaches the deformation is introduced in a rather
ad hoc fashion. In this paper we have provided a novel mechanism, which has
real solutions of physically motivated nonlinear wave equations as the
starting point. The constrained motion of some solitonic solutions of these
models then led to complex Calogero models, some of them being $\mathcal{PT}$%
-symmetric.

There are some obvious open problems left. For instance it would naturally
be very interesting to study systems involving larger numbers of particles,
which would correspond to higher soliton solutions for the nonlinear wave
equations. Clearly the study of different types of wave equations, such as
the KdV etc and their $\mathcal{PT}$-symmetrically deformed versions would
complete the understanding.

Our analysis is schematically summarized in figure 2.

\begin{center}
\includegraphics[width=16cm]{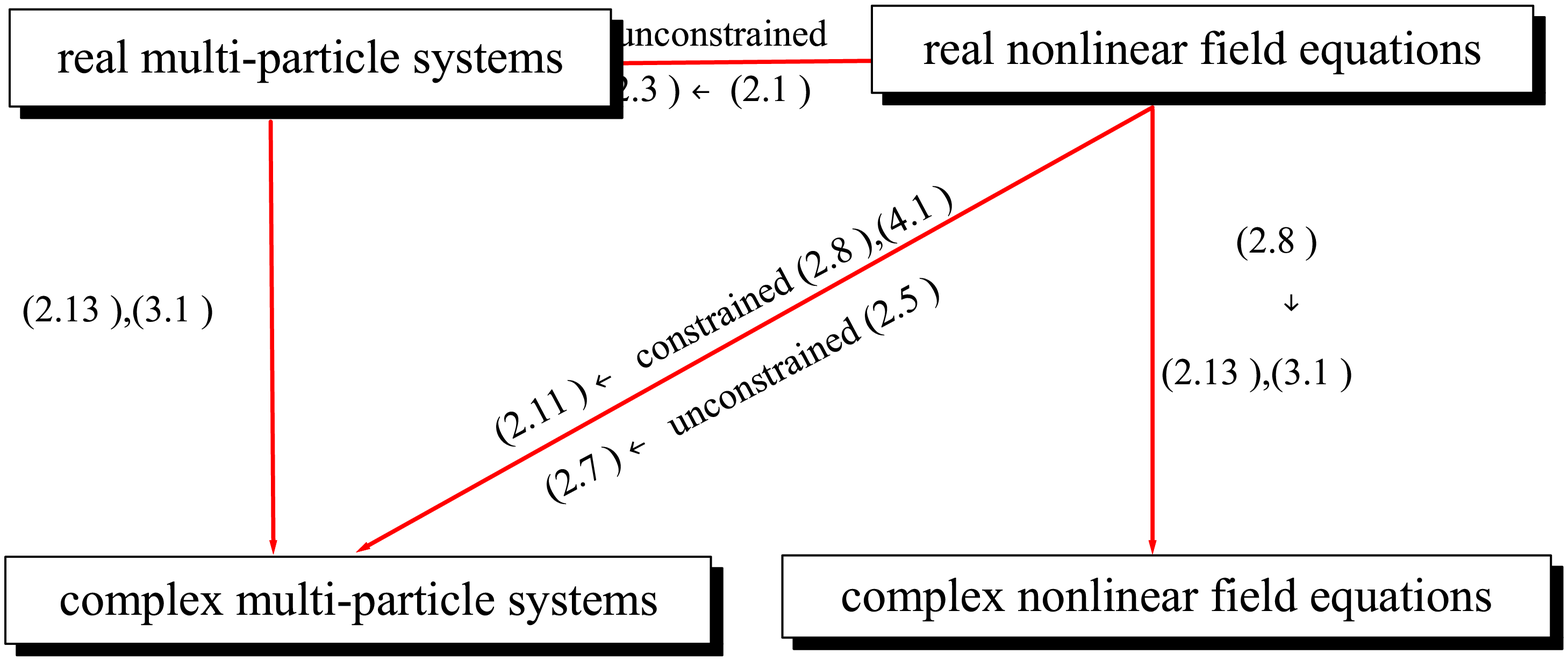}
\end{center}

\medskip

\noindent \textbf{Acknowledgements:} P.E.G.A. is supported by a City
University London studentship. We are grateful to Monique Smith for useful
discussions.

\end{document}